\pdfoutput=1

\documentclass[noshowpacs,amsmath,
preprint,endfloats*,
aps,prb]{revtex4}
\usepackage{graphicx}
\usepackage{amsmath}
\newcommand{\vect}[1]{\textbf{#1}}

\begin{document}

\title{Time-resolved Dynamics of the Spin Hall Effect}

\author{N. P. Stern}
\author{D. W. Steuerman}
\author{S. Mack}
\author{A. C. Gossard}
\author{D. D. Awschalom}
\email[Correspondence and requests for materials should be addressed to D.D.A. (e-mail: awsch@physics.ucsb.edu).]{}
\affiliation{Center for Spintronics and Quantum Computation,
University of California, Santa Barbara, CA 93106 USA}

\date{\today}

\maketitle

\textbf{The generation and manipulation of carrier spin polarization in semiconductors solely by electric fields has garnered significant attention as both an interesting manifestation of spin-orbit physics as well as a valuable capability for potential spintronics devices \cite{Datta:1990, Wolf:2001, Zutic:2004, Kato:2005}.  One realization of these spin-orbit phenomena, the spin Hall effect (SHE) \cite{Dyakonov:1971, Hirsch:1999}, has been studied as a means of all-electrical spin current generation and spin separation in both semiconductor and metallic systems.  Previous measurements of the spin Hall effect\cite{Kato:2004b, Wunderlich:2005, Valenzuela:2006, Sih:2006, Stern:2007} have focused on steady-state generation and time-averaged detection, without directly addressing the accumulation dynamics on the timescale of the spin coherence time.  Here, we demonstrate time-resolved measurement of the dynamics of spin accumulation generated by the extrinsic spin Hall effect in a doped GaAs semiconductor channel.  Using electrically-pumped time-resolved Kerr rotation, we image the accumulation, precession, and decay dynamics near the channel boundary with spatial and temporal resolution and identify multiple evolution time constants.  We model these processes using time-dependent diffusion analysis utilizing both exact and numerical solution techniques and find that the underlying physical spin coherence time differs from the dynamical rates of spin accumulation and decay observed near the sample edges.}

Theories have predicted\cite{Dyakonov:1971, Hirsch:1999, Murakami:2003, Sinova:2004}, and experiments confirmed\cite{Kato:2004b, Valenzuela:2006}, that an electric current in a crystal with spin-orbit coupling gives rise to a transverse spin current via the spin Hall effect.  Spin-dependent scattering of carriers by charged impurities (the extrinsic SHE)\cite{Dyakonov:1971, Hirsch:1999, Engel:2005} or the direct effect of spin-orbit coupling on the band structure (intrinsic SHE)\cite{Murakami:2003, Sinova:2004} causes spin-dependent splitting in momentum space and a resulting pure spin current.  Although not directly observable, the presence of this spin current can be inferred from the existence of non-equilibrium spin accumulation near sample boundaries.  While extrinsic spin Hall currents generated by impurity scattering evolve on momentum scattering timescales ($< 1$ ps), spin Hall accumulation is expected to develop on the much slower spin coherence timescale $\tau$ ($\sim 1$ ns).  Since this timescale is of the same order as that desired for fast electrical manipulation of spin polarization in spintronics devices, understanding dynamics on this timescale is critical for both physical and practical insights to the extrinsic SHE processes.

Steady-state observations of electrically-generated spin accumulation \cite{Kato:2004a, Kato:2004b, Crooker:2005b} are effective for inferring $\tau$, but they cannot directly access the dynamical processes on the nanosecond timescale.  In contrast, time-resolved spin dynamics with picosecond resolution are routinely measured using ultra-fast optical pump-probe techniques\cite{Awschalom:1985, Baumberg:1994}.  Time-resolution of bulk current-induced spin polarization (CISP) was achieved using a photoconductive switch\cite{Kato:2004a}, but only precessional dynamics were observed due to short duration of the ultra-fast current pulse.  Further, in contrast to SHE, CISP is a bulk phenomenon and consequently neither the steady-state\cite{ Kato:2004a, Kato:2005} nor the time-resolved\cite{ Kato:2004a} measurements investigated spatial dynamics near the sample edge.  In the present work, we combine the spatial resolution afforded by scanning Kerr microscopy\cite{Stephens:2003} with an optical probe pulse delayed relative to the electrical pump pulse to achieve both temporal and spatial resolution of spin polarization generated electrically by the SHE in an $n$-doped GaAs channel.  Details of the experimental technique are shown in Figure 1 and are discussed in the Methods section below.

\begin{figure}\includegraphics{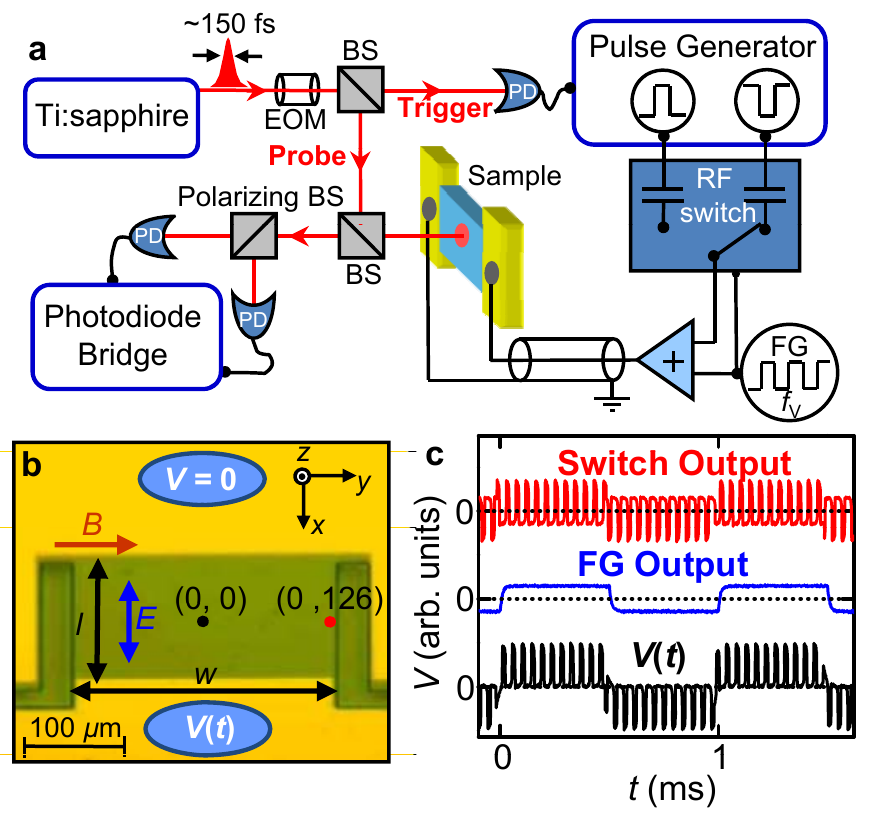}\caption{\label{fig1}
\textbf{Experimental design for time-resolved measurement of electrically-generated spin polarization. a,} Schematic diagram of time-resolved measurement of SHE (EOM, electro-optic modulator, BS, Beam splitter, PD, photodiode, FG, Function Generator).   \textbf{b,} Optical microscope image of the sample with coordinate system defined (units in microns) showing the origin (black circle) and the location for most of the measurements (red circle).  The bright yellow regions are the gold contacts and the grey region is the GaAs channel. \textbf{c} Illustration of the ac pulse scheme for lock-in detection.  The pure ac components of the switch output (red) is added to a square wave at $f_V$ to create a triggered pulse train with amplitude modulation at $f_V \approx 1$ kHz.
}\end{figure}

\begin{figure}\includegraphics{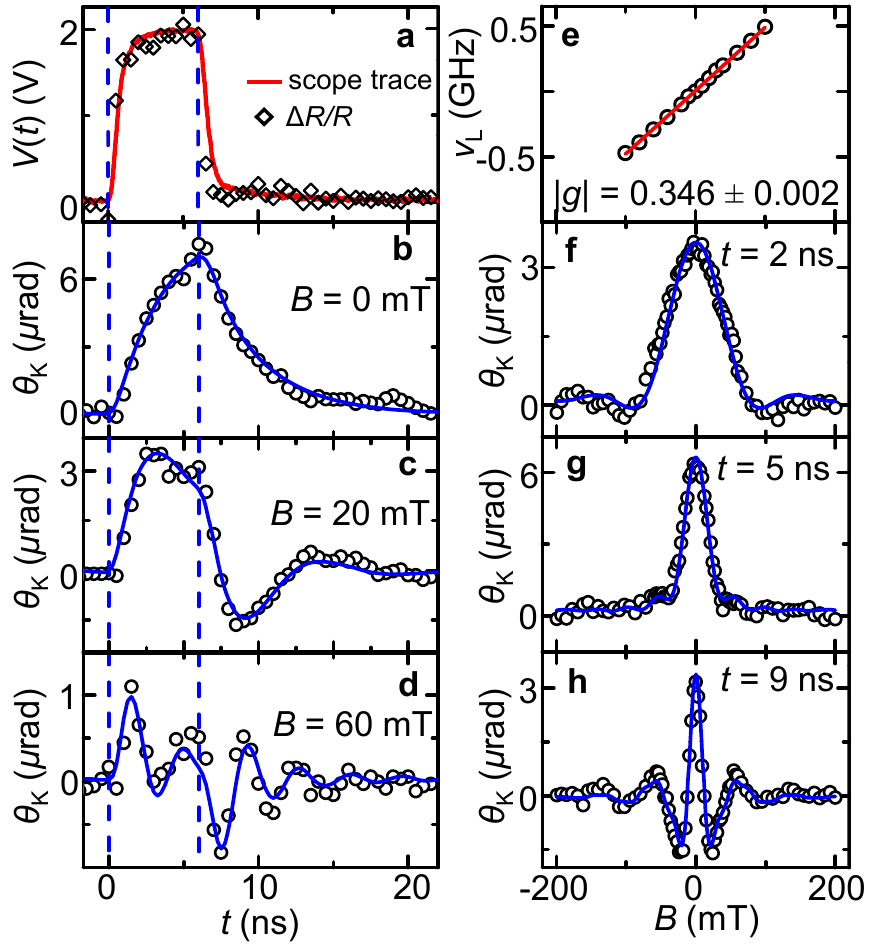}\caption{\label{fig2}
\textbf{Time-resolved measurement of the spin Hall accumulation. a,} Voltage pulse profile $V(t)$ measured by an oscilloscope (red line) and by reflectivity modulation $\Delta R/R$ (black diamonds).  The arbitrary units of $\Delta R/R$ are scaled to match the scope voltage.  The vertical dotted lines mark the time region $0 < t < t_p$. \textbf{b - d,} Representative scans of Kerr rotation $\theta(t)$ for $t_p$ = 6  ns and $y = 126$ $\mu$m.  The blue lines are calculations from Eq. \ref{continuity} with $\tau = $4.2 ns, $L_s = 3.9$ $\mu$m, and $y = 126$ $\mu$m.  (e) $\nu _L$ extracted from cosinusoidal fits to $\theta_K(t)$.  \textbf{f - h,} Kerr rotation $\theta_K(B)$ as a function of magnetic field at three representative times.  The blue lines are calculations from Eq. \ref{continuity}.
}\end{figure}

Figure 2b-d shows the Kerr rotation$\theta_K(t)$ as a function of delay time $t$ in a fixed magnetic field $B$ applied along the $y$-axis.  The spin polarization is generated by the SHE from a voltage pulse $V(t)$  with length $t_p = 6$ ns and amplitude $V_0 = 2$ V (Fig. 2a).  The laser is positioned at $y = 126$ $\mu$m so as to be close to the boundary spin generation with minimal clipping of the spot by the edge.  During the pulse ($0 < t < t_p$), spin polarization builds up due to the SHE.  After the pulse has passed ($t > t_p$), the accumulated spin polarization undergoes decay and spin precession (Fig. 2b-d).  We fit $\theta_K(t)$ for $t > t_p$  to an exponentially decaying cosine to extract an  inhomogeneous depolarization time $\tau^*$ and Larmor precession frequency $\nu_L = g \mu_B B/h$ where $g$ is the Lande g-factor, $\mu_B$ the Bohr magneton, and $h$ is Planck's constant.  Fits to $\nu_L(B)$ give $|g| = 0.346 \pm 0.002$ (Fig. 2e), which is consistent with $g$ expected for this doping level\cite{Yang:1993}.  The depolarization time $\tau^* = 2.8 \pm 0.1$ ns measured at a specific spatial location should differ from the physical spin decoherence time $\tau$.  Because spin polarization can diffuse due to accumulation gradients, there is a second pathway for spin depolarization beyond decoherence which depends strongly on the measurement location.  We reconcile the dynamically measured $\tau^*$ with the intrinsic decoherence time $\tau$ later in our discussion.

Typical optical studies of electrically generated spin accumulation measure the time-averaged projection of spin precession as a function of applied transverse magnetic field $B$.  In analogy with the Hanle effect of depolarization of luminescence, $s_z$ should depolarize for increasing $B$ when $2 \pi \nu_L \sim 1/\tau$.\cite{Engel:2008} Therefore, coherence times in steady-state experiments are typically extracted from the linewidths of $s_z(B)$.  Near the sample edge, $s_z(B)$ is a Lorentzian lineshape analogous to the Hanle effect\cite{Kato:2004b}, whereas it becomes more complicated away from the edges due to the interplay of spin precession and diffusion\cite{Stern:2007, Engel:2008}.  In the current experiment, measurement of $\theta_K(B)$ does not represent a time-averaged steady-state accumulation, but rather a snapshot at a fixed time $t$ of the dynamic behavior of an electrically-generated spin ensemble in a magnetic field.

Representative scans of $\theta_K(B)$ at $y = 126$ $\mu$m are shown for $t = $ 1, 5, and 9 ns in Fig. 2f-h with the magnetic field applied along the $y$ axis.  For small $t$, $\theta_K(B)$ grows in a broad peak that narrows as $t$ increases (Fig. 2f).  Only for $t \sim t_p > \tau$ does $\theta_K(B)$ approach the Lorentzian lineshape expected from a conventional Hanle analysis (Fig. 2g). For $t>t_p$, $\theta_K(B)$ is primarily governed by spin precession, displaying characteristic periodic lobes of decreasing amplitude away from $B = 0$ (Fig. 2h).

In Fig. 3a, we use a longer pulse $t_p = 15$ ns to investigate accumulation dynamics with the current flowing for various $V_0$.  We fit $\theta_K(B=0)$ to an exponential saturation with a time constant $\tau_{acc}$.  For each $V_0$, $\tau_{acc}$ is around 40\%  of the $\tau^*$ measured from decay of the spin polarization (Fig. 3b).   Both $\tau^*$ and $\tau_{acc}$ decrease weakly with $V_0$, which is expected due to electron heating\cite{Stern:2007, Beck:2006}.

\begin{figure}\includegraphics{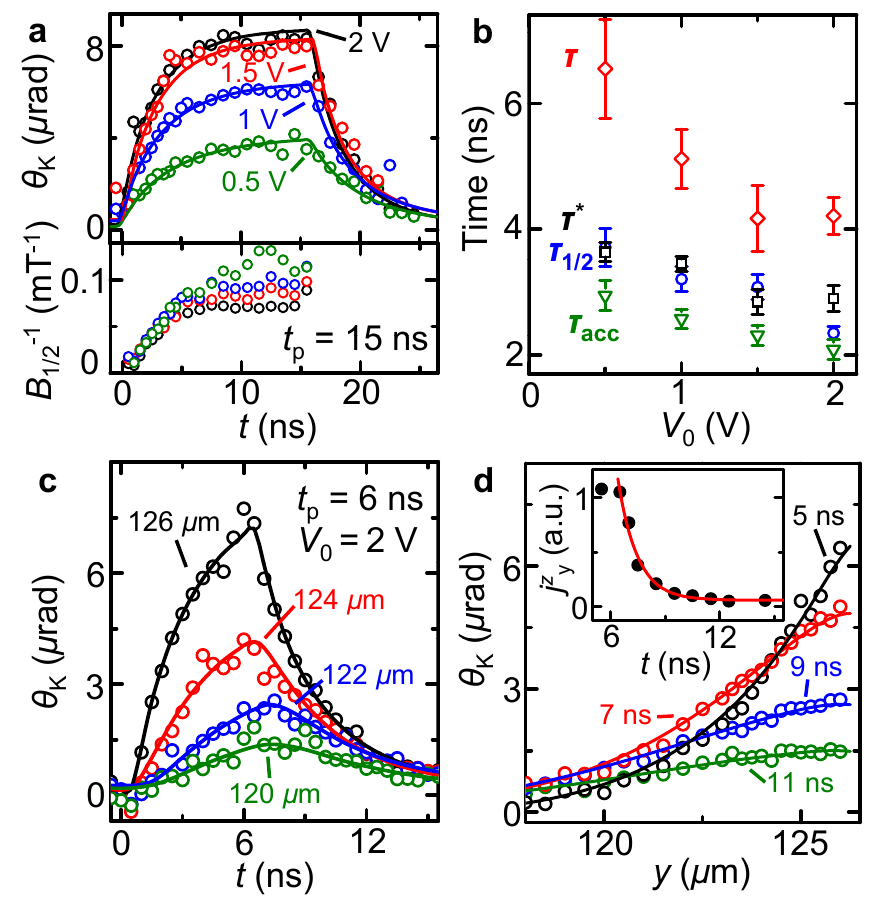}\caption{\label{fig3}
\textbf{Dynamics of spin Hall accumulation. a,} Kerr rotation $\theta_K(t)$ (upper panel), and inverse field width $B_{1/2}^{-1}$ (lower panel) with $t_p = 15$ ns at $y = 126$ $\mu$m for $V_0 = 2.0$ V (black), 1.5 V (red), 1.0 V (blue), and 0.5 V (green).  The solid lines are fits from our model.  \textbf{b,} The accumulation time $\tau_{acc}$ (green triangles), Hanle time $\tau_{1/2}$ (blue circles), decay time $\tau^{*}$ (black squares), and coherence time $\tau$ (red diamonds) as function of voltage.  \textbf{c,} $\theta_K(t)$ for $y = 126$ $\mu$m (black), 124 $\mu$m (red), 122 $\mu$m (blue), and 120 $\mu$m (green).  The solid lines are calculations with amplitude fixed by a fit at $y=126$ $\mu$m. \textbf{d,} Spatially-resolved Kerr rotation $\theta_K(y)$ near the edge for $t = 5$ ns (black), 7 ns (red), 9 ns (blue), and 11 ns (green).  The inset shows the decay of $j^z_y \propto \partial s_z(y)/\partial y$ extracted at $y = 126$ $\mu$m after the voltage pulse is complete ($t_p = 6$ ns).
}\end{figure}

We characterize the magnetic field lineshapes by their inverse half-width $B_{1/2}^{-1}$, which increases with $t$ before quickly saturating (Fig. 3a).  We can understand this evolution of $B_{1/2}$ in a simple physical picture\cite{Engel:2008}.  Soon after the pulse turns on at $t = 0$, spins are all recently generated at the sample edge and have had little time for spin precession about $B$.  For later times, spins have a larger spread in generation times (up to $t_p$) and have correspondingly more time for precession; hence, there is more depolarization for a given $B$ and the Hanle curve narrows (Fig. 2g).  For $t \gg \tau$, the average precession time is governed by $\tau$ rather than $t$ and the Hanle width becomes constant as in a steady-state measurement.  The coherence times $\tau_{1/2}$ calculated from the saturation of $B_{1/2}$ near $t\sim t_p$, agree with decay times $\tau^*$ and are consequently also longer than the accumulation times $\tau_{acc}$ (Fig. 3b).  Diffusion analysis of the SHE accumulation is necessary to reconcile the observed differences between the timescales $\tau_{acc}$, $\tau^*$, and $\tau_{1/2}$.

Electrically-generated spin accumulation in GaAs can be modeled using drift-diffusion equations\cite{ Tse:2005, Engel:2008}.  We treat the channel as infinitely long since its length $l$ is much larger than the spin diffusion length $L_s = 3.9$ $\mu$m found from steady-state measurements at $B=0$.  This assumption reduces the problem to only one spatial dimension and precludes the need for general two-dimensional modeling including spin drift\cite{Stern:2007}.    The SHE generates a spin current transverse to the in-plane electric field $\vect{E}= E\hat{\vect{x}}$ and proportional to the spin Hall conductivity $\sigma_{SH}$, $j^i_j = \sigma_{SH}\epsilon^{ijk} E_k$.    The total current of the $i$ spin component along $y$ from both diffusion and SHE is $j^i_y = -D \partial_y s_i - \sigma_{SH}E\delta_{iz}$, where $D = L_s^{2}/\tau$ is the spin diffusion constant.  Spin conservation at sample edges $y = \pm w/2$ is enforced with hard-wall boundary conditions normal to the edge ($j^i_y = -D \partial_y s^i - \sigma_{SH}E\delta_{iz}=0$).  Diffusive boundary conditions accounting for spin-orbit effects at the sample edge (such as in the intrinsic SHE) would require modification for effects on the scale of the mean free path\cite{Engel:2008}.  Assuming slow spin decoherence (relative to momentum scattering) at rate $1/\tau$, the spin polarization will obey a continuity equation for $\vect{s}(y,t)$ including decay and precession terms:
\begin{equation}\label{continuity}
\frac{\partial s^i}{\partial t}(y, t) = -\frac{\partial j^i_y}{\partial y}(y,t) - \frac{s^i (y,t)}{\tau} + (g \mu_B/\hbar) \vect{B}\times \vect{s}(y,t)
\end{equation}
We first develop an intuitive picture of the time-dependent spin Hall processes using the exact solution to Eq. \ref{continuity} under the simplest conditions.   For $B = 0$, the components of $\vect{s}$ are uncoupled and Eq. \ref{continuity} can be solved using a Green's function to obtain an infinite series solution for $s_z(y, t)$.   The diffusion equation for $s_z$ can be written as:
\begin{align}\label{yDiff}
\frac{\partial s_z}{ \partial t}(y,t) - D \frac{\partial^2 s_z}{\partial y^2}(y,t) + \frac{s_z}{\tau}(y, t) = F(y, t)  \\
F(y, t) \equiv \nabla \cdot (\sigma_{SH} \vect{E}) = \sigma_{SH} \frac{\partial E}{\partial y} \nonumber
\end{align}
$F(y, t)$ is a source function which contains all SHE terms.  The homogeneous $F = 0$ Green's function for Eq. \ref{yDiff} is:
\begin{equation}\label{Greens} G(y, y^\prime, t, t^\prime) =  \sum_m \sin [k_m y] \sin [k_m y^\prime] e^{-\lambda_m(t- t^\prime)} \Theta(t-t^\prime) \end{equation}
where $m$ is an integer, $\Theta(t)$ is the Heaviside step function, $k_m = (2 m +1) \pi/w$ and $\lambda_m = 1/\tau + k_m^2 L_s^2 /\tau $.  For time-independent $E$, integrating Eq. \ref{Greens} yields a series representation of the one-dimensional steady-state solution to the spin Hall diffusion equation.\cite{Kato:2004b} To obtain a time-dependent solution, we assume an ideal square electric field pulse of width $t_p$ and amplitude $E_0$, $E = E_0 [\Theta(y+w/2) - \Theta(y - w/2)] [\Theta(t) - \Theta(t - t_p)]$.  The corresponding source function is $F(y,t) =  \sigma_{SH} E [\delta(y+w/2) - \delta(y - w/2)][ (\Theta(t) - \Theta(t - t_p)]$ and the solution is found by integrating:
\begin{align}\label{Solution}
s_z(y, t) & = 2/w \int_{-\infty}^{\infty} dy^\prime \int_{-\infty}^{\infty}dt^\prime G(y, y^\prime, t, t^\prime) \nonumber \\
& = \frac{4 \sigma_{SH} E_0}{w}  \sum_m \frac{(-1)^m}{\lambda_m } \sin[k_m y] \times T_m(t) \nonumber \\
T_m(t) & \equiv
\begin{cases}
0,  &t < 0 \\
(1 - e^{-\lambda_m t}),  & 0 < t < t_p\\
e^{- \lambda_m (t -t_p)}(1 - e^{-\lambda_m t_p}), & t > t_p
\end{cases}
\end{align}
The three regimes of the term-by-term time-dependence function $T_m(t)$ in Eq. \ref{Solution} can each be observed in Fig. 2b.  The lines in Fig. 3a, represent fits of $\theta_K(t)$ to Eq. \ref{Solution} keeping the first 200 terms in Eq. \ref{Solution} and convoluting the solution with the Gaussian profile of the laser spot.  Since $L_s = 3.9 \pm 0.2$ $\mu$m was found independently from steady-state spatial measurements, the only fit parameters are $\tau$ and an overall amplitude scaling.  The best fit values for the parameter $\tau$ are plotted in Fig. 3b and are significantly longer than the experimentally measured timescales $\tau_{acc}$, $\tau^*$, and $\tau_{1/2}$.

In Fig. 3c, we plot $\theta_K(t)$ and calculations from Eq. \ref{Solution} convoluted with the laser profile for $y = 126$, 124, 122, and 120 $\mu$m for $V_0 = 2$ V using $\tau = 4.2$ ns obtained from the earlier fits.  We fix the amplitude of the calculation from a fit to $y = 126$ $\mu$m, and the remaining curves have no free parameters.  For $y$ away from the edge, $\theta_K(t)$ does not grow exponentially in $t$ and we cannot define the time constant $\tau_{acc}$ as in Fig. 3a.  Comparison of calculations from Eq. \ref{Solution} with and without the spot size averaging reveals that the apparent asymmetry between the growth and decay times $\tau_{acc}$ and $\tau^*$ near the edge is primarily due to spatial averaging of these diffusion profiles over the Gaussian laser spot.  The difference between $\tau^{*}$ and $\tau$ is real, however;  dynamically measured spin polarization near the sample edge evolves with a faster time constant than the underlying spin coherence time.

We can understand the fast evolution of spin polarization from the interplay of diffusion and spin decoherence.  Since polarization gradients cause spins to diffuse away from the sample boundary, spin depolarization must occur faster than decoherence of the electrically-generated spins.  These dynamics are captured in the diffusion analysis of Eq. \ref{Solution} by the fast decay rate $\lambda_m$ of terms with large $m$ in Eq. \ref{Solution}.  Higher $m$ terms are primarily responsible for the discrepancy between the best fit value for $\tau$ and the faster timescales $\tau_{acc}$ and $\tau^*$ observed for spin accumulation and decay in Fig. 3b, but they only contribute significantly to $s_z$ near $ y = w/2$ where all terms are in phase.  In this boundary region, timescales should differ most from the coherence time $\tau$.

We numerically calculate $\partial s_z/\partial y$ at $y= 126$ $\mu$m from spatial scans in Fig. 3d.  The spatial derivative of $s_z(y)$ is proportional to the diffusive spin current and is non-zero at the sample edge in the presence of a compensating spin Hall current for $0 < t <t_p$.  After the spin Hall current disappears at $t=t_p$, $\partial s_z/\partial y$ relaxes with time constant $\tau_{j} = 1\pm 0.1$ ns to satisfy the diffusive $j^z_y = 0$ boundary condition (Fig. 3d, inset).  Since $V(t)$ evolves faster than $\tau^{-1}$, we expect the diffusive spin current at the sample edge to relax faster as well.  Infinitesimally close to the boundary, $\partial s_z/\partial y$ should respond to changes in $V(t)$ as fast as the momentum scattering time.  Taking into account the finite laser spot size, we calculate $\tau_{j} = 1.15$ ns from our model, in agreement with the measured value.

\begin{figure}\includegraphics{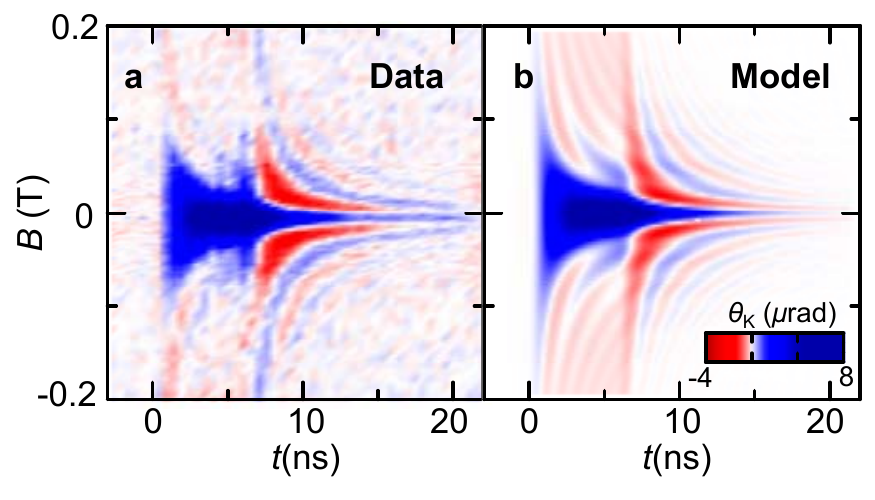}\caption{\label{fig4}
\textbf{Modeling boundary accumulation from the spin Hall effect. a,} Kerr rotation $\theta_K(B, t)$ at $y = 126$ $\mu$m for $t_p$ = 6 ns.  (b) Calculations of $s_z(B, t)$ from Eq. \ref{continuity} with $\tau = 4.2$ and $L_s = 3.9$ $\mu$m at $y = 126$ $\mu$m.
}\end{figure}

Introducing the magnetic field $\vect{B} = B\hat{\vect{y}}$ couples the spin components $s_x$ and $s_z$.  For this regime, we perform numerical solutions to the system of coupled time-dependent linear differential equations represented by Eq. \ref{continuity} using the best fit values for $L_s$ and $\tau$ obtained from the earlier field independent analysis.  For the numerical solutions, we use the exact pulse profile $E(t) = V(t)/l$ measured from the oscilloscope (Fig. 2a) as a source.  The curves in Fig. 2b-d and Fig. 2f-h are numerical calculations of $\theta_K(t)$ and $\theta_K(B)$.  There are no free parameters except an overall scaling to match the amplitude of $\theta_K$.  The full experimental data set $\theta(t, B)$ at $y = 126$ $\mu$m for $t_p = 6$ ns is shown in Fig. 4a.  Figure 4b shows the full calculation of spin accumulation from the numerical solution to Eq. \ref{continuity} for $y = 126$ $\mu$m.

The agreement between the experiments and calculations demonstrates that a single homogeneous decoherence time $\tau$ captures the various timescales observed in time-resolved measurements of the accumulation, decay, and diffusive dynamics of boundary spin polarization due to the extrinsic spin Hall effect.  While diffusive timescales are set by the spin coherence time, evolution near sample boundaries can be limited by the faster response of the spin current.  This spatial dependence of timescales could prove helpful for utilizing electrically-generated spin polarization in high-frequency semiconductor devices.

\section*{Methods}
Channels of width $w$ and length $l$ are processed from a 2-$\mu$m thick silicon-doped GaAs epilayer on 200 nm of undoped Al$_{0.4}$Ga$_{0.6}$As grown on a semi-insulating (001) GaAs substrate by molecular beam epitaxy (Fig. 1b).  The $n$-GaAs has doping density $n = 1 \times 10^{17}$ cm$^{-3}$ and mobility $\mu = 3800$ cm$^2/Vs$ at $T = 30$ K.  The sample is mounted in a helium flow cryostat so that the channel ($x$-direction) is perpendicular to the externally applied in-plane magnetic field $B$ ($y$-direction). A voltage $V(t)$ applied across annealed Ni/Ge/Au/Ni/Au Ohmic contacts creates an in-plane electric field $E(t) = V(t)/l$ along $x$. We desire an impedance of $\sim 50$ $\Omega$ in order to deliver the maximum broadband electrical power to the device; choosing $w = 256$ $\mu$m and $l =130$ $\mu$m yields a device with dc resistance $R = 48$ $\Omega$ at $T = 30$ K.

Time resolution of SHE accumulation is achieved by electrically-pumped Kerr rotation microscopy using a mode-locked Ti:sapphire laser tuned to 1.51 eV that emits a 76-MHz train of $\sim 150$-fs pulses.  The pulse repetition rate is reduced to 38 MHz by pulse picking with an electro-optic modulator.  Each laser pulse is divided into a trigger and a linearly polarized probe pulse.  Spin polarization is generated at the sample edges by the SHE due to the current from a square electrical pulse of width $t_p$, amplitude $V_{0}$, and 0.8 ns rise time applied to the sample from a pulse pattern generator triggered by the optical pump pulse.  The linearly polarized probe beam is focused through a microscope objective to a 1 $\mu$m spot on the surface of the sample which can be scanned with submicron resolution.  A balanced photodiode bridge measures the Kerr rotation of the linear polarization axis $\theta_K$ of the reflected beam which is proportional to the spin polarization along the $z$-axis $s_z$.  The leading edge of the electric pulse profile $V(t)$ arrives at an electronically programmable delay time $t$ before the arrival of the optical pulse. All reported measurements are taken at the center of the length of the channel ($x=0$).

An absorptive RF switch alternates the center conductor of the coaxial cable between the two complementary outputs of the pulse generator at frequency $f_V= 1.337$ kHz.  The RF switch only passes ac components of $V(t)$, so the $0$ V baseline is restored by adding a square wave at frequency $f_V$ back onto the switched pulse train (Fig. 1c) resulting in a modulation of the pulse amplitude $+V_{0}$ and $-V_{0}$ at frequency $f_V$.  $\theta_K$ is then measured with a lock-in amplifier analogous to the ac detection used in Refs. \onlinecite{Kato:2004b, Sih:2006, Stern:2007} but with a definite phase relationship between electrical and optical pulses.

The electrical pulse induces a time-dependent reflectivity modulation $\Delta R/R$ of the optical beam during the pulse duration due to electron heating\cite{Batz:1966, Berglund:1966} that tracks the profile $V(t)$ measured by an oscilloscope (Fig. 2a).  We use this effect to calibrate $t=0$ and confirm that the device acts as a proper 50 $\Omega$ termination due to the minimal temporal pulse distortion at the sample.

\section*{Acknowledgments.}  We thank NSF and ONR for financial support.  N.P.S. acknowledges the support of the Fannie and John Hertz Foundation and S.M. acknowledges support through the NDSEG Fellowship Program.

\section*{}
The authors declare no competing financial interests.


\begin{thebibliography}{25}

\bibitem{Datta:1990} Datta, S. and Das, B. Electronic analog of the electro-optic modulator. \textit{Appl. Phys. Lett.} \textbf{56}, 665-667 (1990).
\bibitem{Wolf:2001} Wolf, S. A. \textit{et al}.  Spintronics: a spin-based electronics vision for the future. \textit{Science} \textbf{294}, 1488-1495 (2001).
\bibitem{Zutic:2004} \v{Z}uti\'{c}, I., Fabian, J., and Das Sarma, S.  Spintronics: Fundamentals and applications. \textit{Rev. Mod. Phys.} \textbf{76}, 323-410 (2004).
\bibitem{Kato:2005} Kato, Y. K., Myers, R. C., Gossard, A. C. and Awschalom, D. D. Electrical initialization and manipulation of electron spins in an L-shaped strained n-InGaAs channel. \textit{Appl. Phys. Lett.} \textbf{87}, 022503 (2005).
\bibitem{Dyakonov:1971} D'yakonov, M. I. and Perel, V. I.  Current-induced spin orientation of electrons in semiconductors. \textit{Phys. Lett.} \textbf{35A}, 459-460 (1971).
\bibitem{Hirsch:1999}  Hirsch, J. E. Spin Hall effect. \textit{Phys. Rev. Lett.} \textbf{83}, 1834-1837 (1999).
\bibitem{Kato:2004b} Kato, Y. K., Myers, R. C., Gossard, A. C. and Awschalom, D. D. Observation of the spin Hall effect in semiconductors. \textit{Science} \textbf{306}, 1910-1913 (2004).
\bibitem{Wunderlich:2005} Wunderlich, J., Kaestner, B., Sinova, J. and Jungwirth, T. Experimental observation of the spin-Hall effect in a two-dimensional spin-orbit coupled semiconductor system. \textit{Phys. Rev. Lett.} \textbf{94}, 047204 (2005).
\bibitem{Valenzuela:2006} Valenzuela, S. O. and Tinkham, M.  Direct electronic measurement of the spin Hall effect. \textit{Nature} \textbf{442}, 176-179 (2006).
\bibitem{Sih:2006} Sih, V., Lau, W. H., Myers, R. C., Horowitz, V. R., Gossard, A. C., and Awschalom, D. D.  Generating spin currents in semiconductors with the spin Hall effect.  \textit{Phys. Rev. Lett.} \textbf{97}, 096605 (2005).
\bibitem{Stern:2007} Stern, N. P., Steuerman, D. W., Mack, S., Gossard, A. C., and Awschalom, D. D.  Drift and diffusion of spins generated by the spin Hall effect. \textit{Appl. Phys. Lett.} \textbf{91}, 062109 (2007).
\bibitem{Murakami:2003} Murakami, S., Nagaosa, N. and Zhang, S. C. Dissipationless quantum spin current at room temperature. \textit{Science} \textbf{301}, 1348-1351 (2003).
\bibitem{Sinova:2004} Sinova, J. \textit{et al.} Universal intrinsic spin Hall effect. \textit{Phys. Rev. Lett.} \textbf{92}, 126603 (2004).
\bibitem{Engel:2005} Engel, H.-A., Halperin, B. I., and Rashba, E. I.  Theory of spin Hall conductivity in \textit{n}-doped GaAs.  \textit{Phys. Rev. Lett.} \textbf{95}, 166605 (2005).
\bibitem{Kato:2004a} Kato, Y. K., Myers, R. C., Gossard, A. C. and Awschalom, D. D. Current-induced spin polarization in strained semiconductors. \textit{Phys. Rev. Lett.} \textbf{93}, 176601 (2004).
\bibitem{Crooker:2005b} Crooker, S. A. \textit{et  al}. Imaging spin transport in lateral ferromagnet/semiconductor structures. \textit{Science} \textbf{301}, 2191-2195 (2005).
\bibitem{Awschalom:1985} Awschalom, D. D., Halbout, J.-M., von Molnar, S., Siegrist, T., and Holtzberg, F.  Dynamic spin organization in dilute magnetic systems.  \textit{Phys. Rev. Lett.} \textbf{55}, 1128-1131 (1985);
\bibitem{Baumberg:1994} Baumberg, J. J. \textit{et al}.  Ultrafast Faraday spectroscopy in magnetic semiconductor quantum structures.  \textit{Phys. Rev. B} \textbf{50}, 7689-7699 (1994).
\bibitem{Stephens:2003} Stephens, \textit{et al}.  Spatial imaging of magnetically patterned nuclear spins in GaAs.  \textit{Phys. Rev. B} \textbf{68}, 041307(R) (2003).
\bibitem{Yang:1993}  Yang, M. J., Wagner, R. J., Shanabrook, B. V., Waterman, J. R., and Moore, W. J.  Spin-resolved cyclotron resonance in InAs quantum wells: a study of the energy-dependent \textit{g} factor. \textit{Phys. Rev. B} \textbf{47}, 6807-6810R (1993).
\bibitem{Engel:2008} Engel, H.-A.  Hanle effet near boundaries: Diffusion-induced lineshape inhomogeneity. \textit{Phys. Rev. B} \textbf{77}, 125302 (2008).
\bibitem{Beck:2006} Beck, M., Metzner, C., Malzer, S., and D\"{o}hler, G. H.  Spin lifetimes and strain-controlled spin precession of drifting electrons in GaAs. \textit{Eurphys. Lett.} \textbf{75}, 597-603 (2006).
\bibitem{Tse:2005} Tse, W., Fabian, J., \v{Z}uti\'{c}, I., and Das Sarma, S.  Spin accumulation in the extrinsic spin Hall effect.  \textit{Phys. Rev. B} \textbf{72}, 241303(R) (2005).
\bibitem{Batz:1966}  Batz, B.  Reflectance modulation at a Germanium surface.  \textit{Solid State Commun.} \textbf{4}, 241-234 (1966).
\bibitem{Berglund:1966}  Berglund, C. N.  Temperature-modulated optical absorption in semiconductors.  \textit{J. Appl. Phys.} \textbf{37}, 3019-3023 (1966).

\end{thebibliography}
\end{document}